\documentclass[oldversion]{aa}

\usepackage{siunitx}
\DeclareSIUnit\um{\micro\meter}
\DeclareSIUnit\Msun{M_{$\odot$}}

\usepackage{url}
\usepackage[tbtags,fleqn]{amsmath}
\usepackage{amssymb}
\usepackage{amsfonts}
\usepackage[varg]{txfonts}
\usepackage{graphicx}
\usepackage{natbib}

\newcommand{\mincir}{\raise
  -2.truept\hbox{\rlap{\hbox{$\sim$}}\raise5.truept \hbox{$<$}\ }}
\newcommand{\magcir}{\raise
  -2.truept\hbox{\rlap{\hbox{$\sim$}}\raise5.truept \hbox{$>$}\ }}


\begin{document}

\title{Molecular clouds have power-law probability distribution
  functions}
\titlerunning{Molecular clouds have power-law PDFs}
\author{Marco Lombardi\inst{1,3}, Jo\~ao Alves\inst{2}, and Charles
  J.~Lada\inst{3}}
\mail{marco.lombardi@unimi.it} \institute{%
  University of Milan, Department of Physics, via Celoria 16, I-20133
  Milan, Italy \and University of Vienna, T\"urkenschanzstrasse 17,
  1180 Vienna, Austria \and Harvard-Smithsonian Center for
  Astrophysics, Mail Stop 72, 60 Garden Street, Cambridge, MA 02138}
\date{Received ***date***; Accepted 
  ***date***}

\abstract{In this Letter we investigate the shape of the probability
  distribution function of column densities (PDF) in molecular clouds.
  Through the use of low-noise, extinction-calibrated
  \textit{Herschel}/\textit{Planck} emission data for eight molecular
  clouds, we demonstrate that, contrary to common belief, the PDFs of
  molecular clouds are not described well by log-normal functions, but
  are instead power laws with exponents close to two and with breaks
  between $A_K \simeq 0.1$ and \SI{0.2}{mag}, close to the CO
  self-shielding limit and not far from the transition between
  molecular and atomic gas.  Additionally, we argue that the intrinsic
  functional form of the PDF cannot be securely determined below $A_K
  \simeq \SI{0.1}{mag}$, limiting our ability to investigate more
  complex models for the shape of the cloud PDF.}  \keywords{ISM:
  clouds, dust, extinction, ISM: structure, Methods: data analysis}

\maketitle

\section{Introduction}
\label{sec:introduction}

In the past couple of decades, the column density probability
distribution function of dark molecular clouds (hereafter PDF) has
received much attention.  The PDF is arguably one of the easiest
quantities to measure (but see below Sect.~2).  Moreover, it is
robustly predicted by many theoretical studies
\citep[e.g.][]{1994ApJ...423..681V, 1997ApJ...474..730P,
  1998ApJ...504..835S, 2010A&A...512A..81F} to follow a log-normal
distribution, and this prediction has been apparently confirmed (at
least up to a few magnitudes of visual extinction) by several
observations (\citealp{2008A&A...489..143L, 2009ApJ...692...91G,
  2010A&A...512A..67L, 2011A&A...535A..16L, 2013ApJ...766L..17S,
  2014arXiv1404.6526A}; but see \citealp{2010MNRAS.408.1089T}).
Finally, departures from log-normality at high densities have been
associated to the star formation activity of molecular clouds (e.g.,
\citealp{2009A&A...508L..35K, 2010A&A...512A..67L,
  2013ApJ...766L..17S}; see also \citealp{2014Sci...344..183K}).

In spite of the profuse efforts to measure the PDF, many of the
observational results obtained so far are lacking a rigorous
discussion of their range of validity and of the possible systematic
effects on them.  Moreover, PDFs obtained from various
observations have been compared to the theoretical log-normal model,
ignoring the limitations of both the observations and the theoretical
predictions.

In this Letter we reconsider the measurements of PDFs and show that
many of the claims made so far do not pass critical scrutiny.  We
highlight a number of observational issues related to the measurements
of the PDF and show that this quantity cannot be robustly measured
below $A_K \sim \SI{0.1}{mag}$.  Using dust emission maps obtained
from \textit{Herschel} and \textit{Planck} data, we show that for $A_K
\gtrsim \SI{0.2}{mag}$, the PDFs of different molecular clouds follow with
good approximation a power law, whose slope in many cases appears to be
close to, or slightly steeper than, $-2$.  Finally, in the range
\SIrange{0.1}{0.2}{mag,} there is a break from the power law at low
extinctions.

\section{Limitations in the measurement of the PDF}
\label{sec:measurement-pdf}

Technically, the PDF is derived as a simple (normalized) histogram of
the column density measurements within some area of the sky that
includes a molecular cloud.  We assume here that the column
density measurements are expressed in terms of $K$-band extinction
$A_K$ and that data are binned in $\log_{10} A_K$ (that is, the PDF
is really a probability distribution for the logarithm base ten of the
extinction). A direct bin in $A_K$ is also possible, and the
associated probability distribution differs from the logarithmic one by
a simple multiplicative term $\propto A_K$.

Molecular clouds are mostly made of molecular hydrogen and helium, two
species which are very difficult to detect at the low temperatures
that characterize these objects.  As a result, the column density of
molecular clouds, from which the PDF is derived, is generally obtained
from different tracers, such as dust (extinction in the optical and
near-infrared and thermal emission in the far-infrared and
submillimeter) or molecules with significant dipole moments (such as
${}^{12}$CO or ${}^{13}$CO).  Each method has different advantages and
limitations that should be understood and taken into consideration
when comparing the observational PDFs with the theoretical
predictions.

In the rest of this Letter, we assume that column density measurements
are obtained through \textit{unbiased} estimators.  In reality,
different techniques suffer from various biases, which will affect
different parts of the PDF.  However, a discussion of the biases
related to column density measurements is beyond the scope of this
letter.  Here, instead, we consider biases arising in the PDFs
from \textit{unbiased} column density measurements.  Specifically,
here we minimize measurement biases by using a combination of
\textit{Herschel} and \textit{Planck}/\textit{IRAS} emission data,
calibrated with a 2MASS/\textsc{Nicest} extinction map (following
\citealp{2014A&A...566A..45L}).

In general, PDFs are affected by four main biases: resolution, noise,
boundaries, and superposition effects, which are the subjects
of the following sections.

\subsection{Resolution and noise biases}
\label{sec:resol-noise-bias}

Each method of probing the column density has a finite resolution and
different noise levels (with the two being inversely proportional to
each other, for a given sensitivity).  The effects of a finite
resolution on the PDF are difficult to predict and quantify, because they
depend on the (unresolved) small-scale structure of the cloud.
Generally, extinction measurements demonstrate that clouds tend to be
relatively smooth at low column densities; i.e., they show only small
local variations in extinction, compared to their denser parts, which
are very uneven, and therefore the most affected by resolution
\citep[see][]{2010A&A...512A..67L}.  Thus, observations with finite
resolutions will often ``move'' mass from the denser parts to the less
dense ones, and the observed PDFs will thus show a lack of dense
material.

The effects of statistical noise can instead be
characterized better: noise acts by smoothing the intrinsic PDF over $A_K$
with a size of the smoothing kernel equal to the average noise level
within each bin.  Therefore, the noise level sets the
resolution in extinction of the PDF.  Depending on the technique used
to derive the cloud column density, the noise level can be constant in
the field or can vary.  For near-infrared (NIR) extinction studies,
the $K$-band extinction measurements toward a star have a typical
error around \SI{0.15}{mag}, and therefore NIR extinction maps have a
fraction of this noise level (because several individual extinction
measurements are averaged within each resolution element): however,
since the use of more stars per resolution element comes at the price
of lower resolution, typical errors on NIR extinction maps are in
the range \SIrange{0.03}{0.10}{mag}.  Moreover, since the density of
background stars decreases in the denser regions of molecular clouds,
the noise of NIR extinction maps increases with the dust column
density.  This level of noise has a significant impact on the
lower end of the PDF (where the signal-to-noise ratio approaches
unity) and makes it virtually impossible to characterize the PDF for
$A_K \lesssim \SI{0.1}{mag}$ with NIR extinction
\citep[see][]{2014A&A...565A..18A}.  The statistical noise of other
tracers, such as dust emission or CO observations, depends on the
depth of the observation and (at least in principle) can be
significantly below $\SI{0.1}{mag}$ (but see below).

On top of statistical noise, many column density tracers are also plagued
by systematic errors.  As mentioned above, we do not consider these,
but it is worth recalling that extinction studies are affected by
unresolved substructures and foreground stars, especially for high
column densities \citep{2009A&A...493..735L}; dust emission maps
suffer from temperature gradients along the line of sight and
inaccuracies in the dust opacity model; and radio observations are plagued
by a very limited dynamic range (which essentially prevents the study
of the PDF).

\subsection{Projection and boundary bias}
\label{sec:bound-proj-bias}

Our view of molecular clouds is confused by projection effects: the
volume probed to derive the PDF is a cone, and intervening material
along the line of sight essentially makes it impossible to probe the
PDFs at low column densities (and, in some cases, close to the
Galactic plane, at medium densities too).\footnote{This is strictly
  true for dust extinction and emission studies.  However, for
  relatively uncrowded regions, CO measurements have some power to
  remove this confusion using velocity information}

As a consequence, molecular cloud boundaries generally are not well
defined in dust emission or extinction maps.  Even for clouds
relatively distant from the galactic plane (such as Orion, Taurus, or
Perseus) it is difficult to go below $A_K \sim \SI{0.1}{mag}$: that
is, iso-contours corresponding to lower values of extinction of one
cloud are generally merged with unrelated cloud material in the
foreground and background.

Operationally, the choice of the sky area used to derive the PDF
clearly affects the measurement of the PDF. Including or excluding
regions angularly close to the cloud has an impact on the overall
shape of the PDF, especially at low column densities.  For example,
larger boundaries generally tend to extend the PDF to lower values of
$A_K$.

\section{\textit{Herschel}-derived PDF of nearby clouds}
\label{sec:top-end-pdf}

Because of the effects discussed in Sect.~\ref{sec:measurement-pdf},
in order to measure the PDFs of molecular clouds we need to use
well-calibrated data with the highest dynamic range and a large areal
coverage of the clouds.  Therefore, we follow
\citet{2014A&A...566A..45L} by using \textit{Herschel} emission maps
complemented with \textit{Planck}/\textit{IRAS} data for the outskirts
of the clouds to derive column densities.  We finally convert the
optical depth to extinction using 2MASS/\textsc{Nicest} maps.

\begin{figure}[t!]
  \centering
  \includegraphics[width=\hsize]{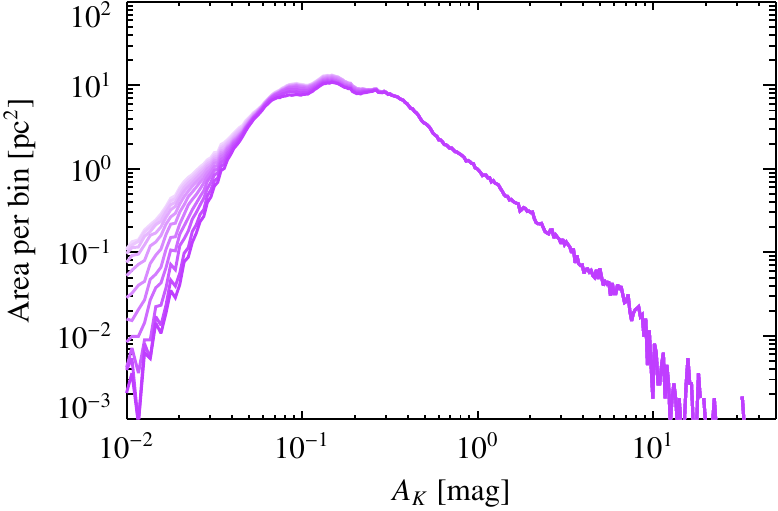}
  \caption{Effect of 11 different boundaries used to derive the
    PDF of Orion~B (dark: smaller area, light: larger area, by equal
    steps of $\sim 4\%$).  The effect on the PDF is almost exclusively
    confined to $A_K < \SI{0.1}{mag}$.}
  \label{fig:1}
\end{figure}

\begin{figure}[t!]
  \centering
  \includegraphics[width=\hsize]{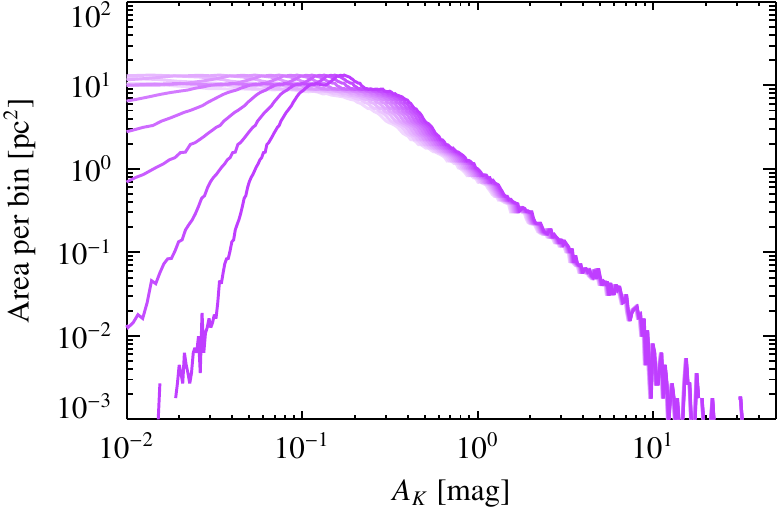}
  \caption{Effect of 11 different offsets for the superposition
    bias correction in the PDF of Orion~B (by steps of
    \SI{0.02}{mag}).}
  \label{fig:2}
\end{figure}

\begin{figure}[t!]
  \centering
  \includegraphics[width=\hsize]{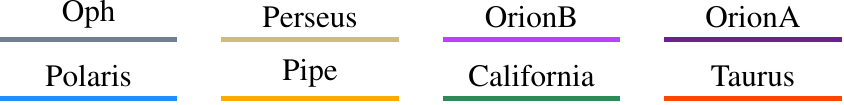}
  \includegraphics[width=\hsize]{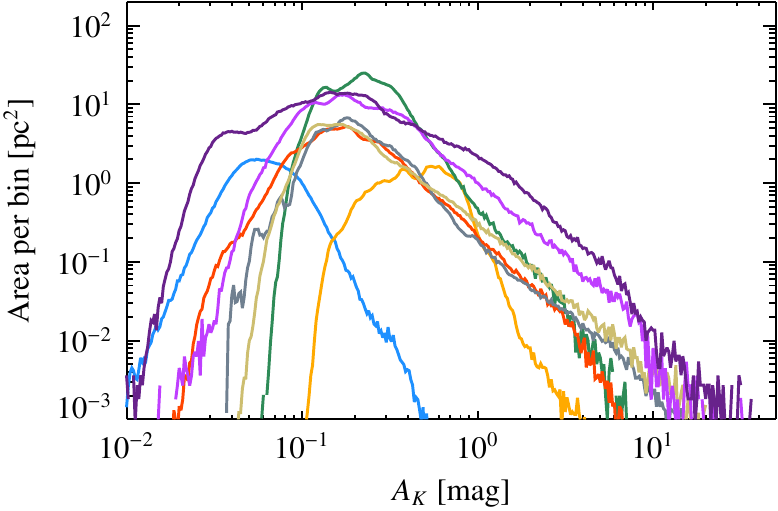}
  \caption{Areas per extinction bin of different molecular clouds
    from \textit{Herschel}/\textit{Planck} dust-emission data. Bins
    span \SI{0.01}{dex}.}
  \label{fig:3}
\end{figure}

\begin{figure}[t!]
  \centering
  \includegraphics[width=\hsize]{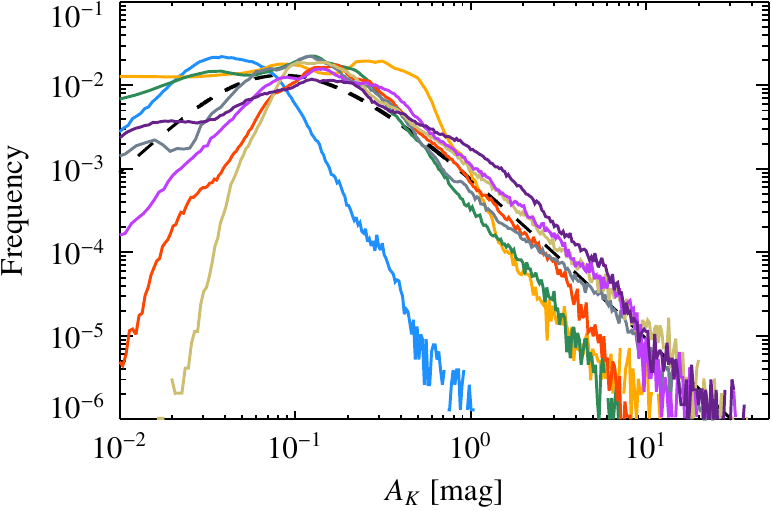}
  \caption{PDFs of molecular clouds considered in this Letter,
    together with the PDF generated by a simple toy model (truncated
    isothermal profile, see Sect.~\ref{sec:discussion}) as a black
    dashed line.  The PDFs have been corrected for the superposition
    bias by subtracting a constant offset to the dust extinction maps
    used to derive them.}
  \label{fig:4}
\end{figure}

As argued in the previous section, the PDF is expected to be affected
by choice of cloud boundaries.  Figure~\ref{fig:1} shows how the
histogram of the bin areas (thus essentially unnormalized PDFs) of
Orion~B changes when using different boundaries.  As expected, this
has a strong impact on $A_K < \SI{0.1}{mag}$, while the high end of
the PDF is left unchanged.  

As mentioned earlier, unrelated foreground or background material can
contribute to the observed PDF.  One way to correct for this is to
look at the lowest extinction value in a large area around the cloud
and to remove this amount from the extinction map (see also
\citealp{2014arXiv1403.2996S}).  Of course, this is a crude
approximation since the subtracted column density is taken to be
constant within the field.  As a result, we expect ``corrected''
column densities to be affected by an additional noise equal to the
average scatter of the superimposed material.  This quantity, however,
can be estimated (although approximately) by checking the off-field
column density scatter and by applying a set of offsets that spans the
same range in extinction.  To test the bias associated with such a
correction, we subtracted different extinction offsets to the PDF of
Orion~B.  The result of this experiment (Fig.~\ref{fig:2})
demonstrates that this operation mostly affects the low end of the
PDF: in particular, large offset corrections make the PDF peak broader
(in a log-log plot) and move it to the left.

\begin{table}
  \centering
  \tabcolsep=0.15cm
  \begin{tabular}{lccc@{\hspace{0.7cm}}lccc}
    Cloud   & $\Delta A_K$ & $n$ & $b$ &
    Cloud   & $\Delta A_K$ & $n$ & $b$ \\
    \hline
    Oph     & 0.06 & 1.8 & $\phantom{-}17^\circ$ &
    Perseus & 0.02 & 1.7 & $-20^\circ$ \\ 
    Orion~B & 0.03 & 2.0 & $-15^\circ$ &
    Orion~A & 0.02 & 1.9 & $-19^\circ$ \\
    Polaris & 0.01 & 3.9 & $+25^\circ$ &
    Pipe    & 0.29 & 3.0 & $\phantom{-0}5^\circ$ \\
    California & 0.10 & 2.5 & $\phantom{0}{-8^\circ}$ & 
    Taurus     & 0.01 & 2.3 & $-15^\circ$ 
  \end{tabular}
  \caption{Extinction correction, the computed slopes $n$ of the
    power law of the various clouds' PDFs and the clouds'
    galactic latitudes $b$.}
  \label{tab:1}
\end{table}

These simple tests demonstrate that the low end of the PDF is
essentially unconstrained by the observations.  We thus limit
our investigation to the PDF at medium-to-high column densities.
Figure~\ref{fig:3} shows the raw histograms of bin areas for a set of
molecular clouds, with boundaries selected from 2MASS/\textsc{Nicest}
extinction maps (see \citealp{2006A&A...454..781L,
  2008A&A...489..143L, 2010A&A...512A..67L,
  2011A&A...535A..16L}).\footnote{As discussed in the text, this
  figure is constructed directly from histograms of the logarithm of
  the column density map of each cloud, and differs by a simple
  scaling factor $\propto A_K$ from Figs.~17 and 18 of
  \citet{2014A&A...566A..45L}, which are constructed as derivative of
  the area function.}\@.  We stress that using \textit{Planck}
data for the outskirts of the clouds was critical for investigating
regions outside the \textit{Herschel} coverage but still at relatively
high values of extinction.  Clearly there is a wide variety of PDF
shapes, and in almost all cases there, they do not look like simple
log-normal functions (which would appear as parabolae here).  However,
as discussed earlier, each cloud is affected by different levels of
contamination due to unrelated foreground and background material.  To remove this bias, we proceed as in Fig.~\ref{fig:2} and
subtract, for each cloud, a custom offset appropriately determined by
careful examination of the outskirts of each object and list in
Table\ref{tab:1}.

The result is shown as (normalized) PDFs in Fig.~\ref{fig:4}.
Interestingly, many of the differences at $A_K \sim
{}$\SIrange{0.1}{0.5}{mag} evident in Fig.~\ref{fig:3} are absent or
mitigated in Fig.~\ref{fig:4}, suggesting that they are artificially
induced by superposition of unrelated foreground and background
material.

At extinctions $> \SI{0.2}{mag}$, the PDFs exhibit power-law shapes
spanning approximately two decades with slopes ranging roughly between
$-4$ and $-2$ (see Fig.~\ref{fig:4} and Table\ref{tab:1} where we list
the indexes derived from a fit of the PDFs).  In addition, they exhibit
a turnover from a power law form near $A_K \sim \SI{0.2}{mag}$.
Log-normal PDFs, which would appear as simple symmetric parabolae in
these plots, are not evident for $A_K > \SI{0.1}{mag}$, with the
possible exception of Polaris.  Rather, it is evident that PDFs are
very asymmetric in the log-log plot.  Again, Polaris is a notable
exception in this plot: it is more symmetric and displays a break at
significantly lower column densities, as well as a steeper slope at the
high extinction side.  As argued above, the differences shown by
clouds below $A_K \sim \SI{0.1}{mag}$ are not significant;
therefore, we cannot even assess whether there is a universal shape for the
PDF in this regime, and in case there is, if the PDF is flat or
increasing in this interval.  Moreover, in this regime much of the
measured extinction is likely to come from unrelated, more
diffuse, atomic and molecular gas along the line of sight, rather than
from the molecular cloud itself.  At such low extinctions, it would be
very difficult to separate any contribution to the intrinsic cloud PDF
arising from a cloud's own (more diffuse) outermost layers.

\section{Discussion}
\label{sec:discussion}

Our inability to investigate the low end of the PDF limits our ability
to distinguish different models for its shape.  However, the data
discussed in this letter already provides enough information to draw
some conclusions.

First, there is no indication that these PDFs are simple log-normal
functions.  Second, at high extinction, the PDFs are best described by
power laws, with a break at $A_K \sim \SI{0.15}{mag}$.  We observe
that clouds with lower star formation activity (such as Polaris and
Pipe) seem to be characterized by steeper slopes than more active star-forming clouds (such as Orion~A and B, Ophiuchus, and Perseus). This
is expected, since a steeper slope implies a lack of
high-density material, hence reduced star formation activity (see
\citealp{2013A&A...559A..90L} and \citealp{2013ApJ...778..133L}).
Interestingly, this result is consistent with recent simulations of
turbulent cloud evolution that suggest that the slopes of the high
extinction end of PDFs systematically vary with the age of the
molecular cloud \citep{2014MNRAS.445.1575W}.  In these simulations,
\citeauthor{2014MNRAS.445.1575W} find that an initially log-normal PDF
develops a power-law tail that becomes increasingly prominent until it
ultimately dominates the PDF above extinctions of $A_K \sim
\SI{0.2}{mag}$.  During this evolution the power-law indices
systematically decrease with time, approaching a value of $-2$ after
about \SI{5}{Myr}.  Although it is true that some portion of the PDF
above \SI{0.2}{mag} could still be fit by the arc of a broad
log normal with a peak at lower extinctions
\citep{2014arXiv1404.6526A,2014MNRAS.445.1575W}, the observations
simply require a function no more complicated than a power law in this
regime.  Unfortunately, the various biases present prohibit our
ability to investigate the actual form of the cloud PDF below $\sim
\SI{0.1}{mag.}$  Even though more complex models might account for,
and be consistent with, a portion of the PDF above \SI{0.2}{mag}, they
are not required by the data.  Thus, although other observations, such as
cloud velocity fields, fractal boundaries, and scaling relations, may
physically motivate turbulent cloud models, the observed PDFs by
themselves cannot because in the regime where a log-normal
distribution might exist, the intrinsic PDFs cannot be probed
\citep[see also][]{2012MNRAS.423.2579B}.

There is a clear indication that the power-law regime has a break at
low values of extinction.  Since the break is in an area that is still
largely unaffected by the biases discussed in this Letter (see
Fig.~\ref{fig:1}), we trust that the break is real.  The location of
the break coincides approximately with the column density required for
CO self-shielding and is near the column density threshold for the
$\mathrm{H}_2$-to-HI transition ($A_K \sim 0.04$ to \SI{0.1}{mag};
\citealp{2014ApJ...790...10S, 2011A&A...536A..19P,
  2010ApJ...716.1191W}).  This suggests that the break might be
related to either or both of these thresholds. Also, the slope $\sim
-2$ of the power law is reminiscent of an isothermal profile
\citep{2014A&A...566A..45L}. Therefore, it is reasonable to consider
the PDF associated with a pressure-truncated isothermal profile to see
if the associated PDF shares any similarity with the PDFs observed in
real molecular clouds.  We consider the toy model of truncated
(singular) isothermal profile.  The associated PDF can be provided in
parametric form:
\begin{align}
  \label{eq:1}
  \Sigma(r) & {} \propto f(r) = \frac{2}{r} \arctan \left( \frac{\sqrt{1 -
        r^2}}{r}
  \right) \; , \\
  \label{eq:2}
  \mathit{PDF}(r) & {} \propto - 2 r \frac{f(r)}{f’(r)} \; .
\end{align}
In this parametrization, $r$ is the radius normalized to the
truncation radius and $\Sigma(r)$ the column density or extinction.
In the limit of small $r$, substituting Eq.~\eqref{eq:1} in
Eq.~\eqref{eq:2}, we obtain $\mathit{PDF} \propto \Sigma^{-2}$.  By
plotting Eqs.~\eqref{eq:1} and \eqref{eq:2}, one sees that the
function implicitly defined here broadly resembles the PDFs of many
molecular clouds considered in this Letter (Fig.~\ref{fig:4}): this
function has a slope of $-2$ for high column densities, and it peaks
at the radius $r_\mathrm{break} \simeq 0.72 r_\mathrm{trunc}$, i.e.,\
close to the truncation radius.

Of course, it is unrealistic to think that real clouds can follow
spherically symmetric singular isothermal profiles exactly, with
perfectly sharp outer edges.  More naturally, an effective pressure
truncation could result from a rapid increase in the local cloud sound
speed, owing for example, to the transition from molecular to atomic gas
or to a rapid increase in temperature or turbulence across the cloud
boundaries.  In the spherical approximation, $A_K \propto \Sigma
\propto r^{-1}$ and the break of the PDF occurs at $1/0.72 \simeq 1.4$
times the truncation column density.  Since we observe the peak in the
PDFs of many clouds at $A_K \sim {}$\SIrange{0.1}{0.2}{mag}, we can
argue that the truncation must occur around $A_K \sim
{}$\SIrange{0.07}{0.14}{mag}.  As mentioned above, these values are
not too far from the column density threshold for the
$\mathrm{H}_2$-to-HI transition. (The exact location of this transition
depends on several physical conditions of the cloud, such as the
ultraviolet background and the exposure to cosmic rays.)

Another alternative is to imagine that molecular clouds can be
(approximately) described as the sum of several isothermal spheres
(for example, each corresponding a core).  Individually, the PDF of
each core would be a power law, and therefore the PDF of the
entire cloud would also follow a power law.  However, since the volume
available for each core is limited (by the presence of the other
cores), the resulting PDF shows a depression at low column densities
with respect to the pure power law implied by (infinite) isothermal
profiles.  Therefore, this simple model could qualitatively explain
the general shapes of the observed PDFs.

These interpretations are just a few of the several possible ones.
Unfortunately, our inability to investigate the low end of the PDF
makes it very difficult to distinguish among them.  In particular,
because of the intrinsic limitations of the column density
measurements, presumed log-normal PDFs of clouds cannot be validated
by such observations.

\begin{acknowledgements}
  M.L. acknowledges financial support from PRIN MIUR 2010--2011,
  project ``The Chemical and Dynamical Evolution of the Milky Way and
  Local Group Galaxies.''
\end{acknowledgements}

\bibliographystyle{aa} 
\bibliography{../dark-refs}

\end{document}